# Neutrino propagation in matter and CP violation

Paul M. Fishbane[*]
*Physics Dept. and Institute for Nuclear and Particle Physics,
Univ. of Virginia, Charlottesville, VA 22904-4714*

Peter Kaus[**]
*Aspen Center for Physics, Aspen, CO 81611*

We point out that the dependence on the order of the matter through which neutrinos pass can provide a window into CP violation in the neutrino sector. This allows for study of CP in the neutrino sector without the necessity of making a comparison between the behavior of neutrinos and that of antineutrinos.

[*] e-mail address pmf2r@virginia.edu
[**] e-mail address pkaus@futureone.com

The presence of neutrino oscillations [1-2] leads to the question of possible CP violation in a (fully-coupled) three-family neutrino sector [3]. We suggest here a probe of CP that uses an effect of the modification of neutrino evolution within matter[1] [4-6]. The early two-family work of Refs. 4 and 5 arrives at a clear analytic result for propagation of neutrinos within a medium of constant density. Our method is also clearly illustrated in the two-family case. After development of our method we illustrate the three-family case with some numerical calculations.

Neutrino propagation in material differs from propagation in vacuum because, in contrast to $\nu_\mu$ and $\nu_\tau$, any $\nu_e$ component in the neutrino beam can scatter from electrons in the material through charged current interactions. Consider neutrino passage through a set of $N$ constant density material layers that we label 1, 2,…, $N$. These could be materials of finite width or could be a set of labels that represent an (arbitrarily good) approximation to a material of continuously varying density. Because of multi-family transitions, the order of the materials matters. With $n$ neutrino families, the $n \times n$ amplitude $A^{12...N}$ is

$$\psi(t) = A^{12...N}(t)\psi(0). \tag{1}$$

where $\psi$ are the flavor states and $t$ is the total time (or interchangably the total distance). The multiple-layer amplitude is the ordered product of the amplitudes for passage through the respective single layers,

$$A^{12...N} = A^1 A^2 ... A^N \tag{2}$$

While Eq. (2) is general, to proceed further we want to start with two families, for which we know the single-layer amplitude $A^j$ is symmetric (see below). Then the amplitude for passage through the "reversed" material[2], in which the order of layers is inverted, is

$$A^{reverse} = A^{N...21} = A^N \cdots A^2 A^1 = \left(A^1 A^2 \cdots A^N\right)^T$$
$$= \left(A^{12...N}\right)^T = \left(A^{direct}\right)^T. \tag{3}$$

To see the consequences of this, recall that for two families with mass eigenvalues $m_i^2$ and a single (positive) mass difference factor $\delta m^2 = m_2^2 - m_1^2$, and a Cabibbo matrix describing the flavor states in terms of the energy eigenstates in the usual way, the single-layer amplitude [4, 5] takes the *general* form

$$\begin{bmatrix} \gamma & \eta \\ -\eta^* & \gamma^* \end{bmatrix}. \tag{4}$$

(In fact $\eta$ is pure imaginary, but that is unimportant here.) We can then show by recurrence that the $N$ layer amplitude has the structure

$$A^{direct} = A^{12...N} = \begin{bmatrix} \alpha & \beta \\ -\beta^* & \alpha^* \end{bmatrix}. \tag{7}$$

This follows because the amplitude for $N$ + 1 layers is

$$A^{12...N+1} = A^{12...N} A^{N+1} = \begin{bmatrix} \alpha & \beta \\ -\beta^* & \alpha^* \end{bmatrix} \begin{bmatrix} \gamma & \eta \\ -\eta^* & \gamma^* \end{bmatrix}$$
$$= \begin{bmatrix} \alpha\gamma - \beta\eta^* & \alpha\eta + \beta\gamma^* \\ -\alpha^*\eta^* - \beta^*\gamma & \alpha^*\gamma^* - \beta^*\eta \end{bmatrix},$$

and this indeed has the structure of (7). It remains only to show by direct calculation starting from (4) that a two layer amplitude has the structure of (7), and this is easily done. This result can alternatively be derived in a continuum limit, for which the amplitude has an ordered exponential form.

---

[1] In other work [3], the effects of matter imitate CP violation and it is necessary to distinguish the effects of matter from those of the presence of a CP-violating phase.

[2] We generally refer to quantities for the order 12…N as the "direct" case, and quantities for the order N…21 as the "reverse" case.

The combination of Eqs. (3) and (7) then tells us that

$$A^{reverse} = \begin{bmatrix} \alpha & -\beta^* \\ \beta & \alpha^* \end{bmatrix}. \tag{8}$$

Comparison between Eqs. (7) and (8) reveals that the probabilities $P_{ij} \equiv |A_{ij}|^2$ satisfy
$$P_{ij}^{reverse} = P_{ij}^{direct},$$
although only the diagonal elements of the amplitudes are equal.

The presence of three families complicates the situation. Even if we could establish that Eq. (3) were valid, it would give only a useful relation for the diagonal matrix elements of the direct and reverse amplitudes; assuming the validity of (3) we find

$$A_{jj}^{reverse} = A_{jj}^{direct}. \tag{9}$$

With this statement alone and unitarity, one can deduce the fact that for two families the off-diagonal probabilites are the same; for example,

$$P_{12}^{reverse} = 1 - P_{11}^{reverse} = 1 - P_{11}^{direct} = P_{12}^{direct}.$$

For three families the off-diagonal probabilities for the direct and reverse probabilities would not generally be equal. Rather, for example,

$$P_{12}^{reverse} + P_{13}^{reverse} = 1 - P_{11}^{reverse} = 1 - P_{11}^{direct} = P_{12}^{direct} + P_{13}^{direct}.$$

Let us turn now to the question of three families, including the possibility that the mixing matrix is complex (CP in the mixing matrix). We begin by recalling the amplitude in vacuo. With the CKM matrix $V$ connecting the flavor eigenstate $\psi$ and energy eigenstate $\xi$ according to the $\psi = V\xi$, one can easily show that

$$A(t) = V \begin{bmatrix} e^{-iE_1 t} & 0 & 0 \\ 0 & e^{-iE_2 t} & 0 \\ 0 & 0 & e^{-iE_3 t} \end{bmatrix} V^\dagger$$

(Given the freedom to remove an overall phase, it is possible to use, for example, $E_i \cong E + m_i^2/(2E)$, and pull out factors so that the central matrix gives only dependence on the two mass difference factors $\delta m_{21}^2 \equiv m_2^2 - m_1^2$ and $\delta m_{31}^2 \equiv m_3^2 - m_1^2$ of the problem.) Only if there is no CP violation, so that $V$ is real, is $A$ symmetric. When matter is involved, and ignoring terms proportional to the unit matrix, the problem comes down to diagonalizing the matrix

$$H_{flavor} \equiv \frac{1}{2E} V \begin{bmatrix} 0 & 0 & 0 \\ 0 & \delta m_{21}^2 & 0 \\ 0 & 0 & \delta m_{31}^2 \end{bmatrix} V^\dagger + \begin{bmatrix} V_{cc} & 0 & 0 \\ 0 & 0 & 0 \\ 0 & 0 & 0 \end{bmatrix}. \tag{10}$$

This means finding a unitary matrix U such that

$$UH_{flavor}U^{\dagger} = \begin{bmatrix} \lambda_1 & 0 & 0 \\ 0 & \lambda_2 & 0 \\ 0 & 0 & \lambda_3 \end{bmatrix} \quad (11)$$

The time evolution equation is then solved in the diagonal basis, and after transformation back to the flavor basis we have

$$A(t) = U^{\dagger} \begin{bmatrix} e^{-i\lambda_1 t} & 0 & 0 \\ 0 & e^{-i\lambda_2 t} & 0 \\ 0 & 0 & e^{-i\lambda_3 t} \end{bmatrix} U \equiv U^{\dagger} D(t) U \quad (12)$$

(We remark here that if the consequences of a nonzero $\delta$ are not to be lost, $U$ cannot be found by using approximations based on a presumed neutrino mass hierarchy. Doing so leads to a decoupling in which the oscillation proceeds essentially through two-family steps, and this removes any effect of CP violation [7]. Similarly, if any $\delta m_{ij}^2 = 0$, then $\delta$ can simply be set to 0.)

At this point we can understand the effect of a CP-violating phase $\delta$ in $V$. If $\delta = 0$, then $V$ is real, and hence so is $U$, and $A$ is symmetric. Equation (3) holds for the three channel case, and the respective diagonal elements of the direct and reverse processes are equal to one another. If, however, $\delta$ is non-zero, then $V$ is no longer real, and neither is $U$. More specifically, we can generally write our primitive amplitude (12) with an index $j$ for the material (including its width—$t \to t_j$). Then for passage through a sequence of layers we have

$$\begin{aligned}\left(A^{reverse}\right)^T &= \left(\left(U_N^{\dagger} D_N(t_N) U_N\right) \cdots \left(U_2^{\dagger} D_2(t_2) U_2\right)\left(U_1^{\dagger} D_1(t_1) U_1\right)\right)^T \\ &= \left(U_1^{\dagger}\right)^* D_1^*(-t_1) U_1^* \left(U_2^{\dagger}\right)^* D_2^*(-t_2) U_2^* \cdots \left(U_N^{\dagger}\right)^* D_N^*(-t_N) U_N^* \\ &= \left(A^{direct}(t_i \to -t_i)\right)^*.\end{aligned}$$

In other words, the more general form of Eq. (3) is

$$A^{reverse} = \left(\left(A^{direct}(t_i \to -t_i)\right)^*\right)^T = \left(A^{direct}(t_i \to -t_i)\right)^{\dagger} \quad (13)$$

In particular, the diagonal matrix elements of the direct and reverse processes are no longer equal, and this provides a way to test for the presence of a nonzero value of $\delta$.

Of the diagonal processes available to us, we cannot use the process $\nu_e \to \nu_e$. That is because, given the fact that the charged current occurs in the electron sector, there is no CP violation visible in $\nu_e \to \nu_e$ to leading order in the weak interaction [8]. We have verified the argument with, among others, direct numerical calculation with the parameters described below: the process $\nu_e \to \nu_e$ remains independent of material order and indeed independent of $\delta$.

The process $\nu_\mu \to \nu_\mu$ does allow us to test the effect described here. To illustrate we use a set of numbers that correspond to the so-called large-angle solar MSW scenario [9], namely $\delta m_{32}^2 \cong \delta m_{31}^2 = 3.5 \times 10^{-3}$ eV$^2$, $\delta m_{21}^2 = 5 \times 10^{-5}$ eV$^2$, $\sin\theta_{13} = 0.10$, $\sin\theta_{23} = 0.71$, $\sin\theta_{12} = 0.53$, with two layers, layer 1 corresponding to a material with density 8 gm/cm$^3$ ($V_{cc} = 2.4 \times 10^{-13}$ eV) and layer 2 being that of the vacuum (or, for our purposes, air). We consider three energies, $E_1 = 100$ MeV, $E_2 = 500$ MeV (corresponding to $\delta m^2/(2E) \cong V_{cc}$ in material 1 for $\delta m^2 = 10^{-4}$ eV$^2$), and $E_3 = 20$ GeV (corresponding to a possible neutrino factory energy [3]). In Figs. 1a-c we plot for each energy the difference in probabilities for $\nu_\mu \to \nu_\mu$ for the direct and reverse process. In the range from $\delta = 0$ to $\delta = 1$ rad, the $\delta$ dependence is essentially linear in the magnitude of the differences, with no effect for $\delta = 0$ to the largest effect for the largest $\delta$. Aside from the height magnification there is no additional $\delta$-dependence, and we restrict the plots presented here to $\delta = 0.5$. The horizontal axis is the thickness of the air layer in units of $10^{13}$ eV$^{-1}$; we have in Figs. 1a-c taken equal amounts of vacuum and material.

A well-known characteristic of the types of effects we are talking about is that they are compressed at low energies and stretched out at higher energies, and this is clearly visible in the three figures. We note the following: At $E_1$, the effect tends to be more symmetric about the horizontal axis. Because in any hypothetical experiment the fine oscillations tend to be averaged over, the symmetry is a disadvantage. If one spreads the horizontal axis over a much larger range, say from 0 to 100, then the envelope of the fine oscillations, which occurs in a series of "lobes" as is most visible in Fig. 1b, moves slowly to either side of the horizontal axis, but clearly $E_1$ is not the most interesting case. That is reserved for $E_2$, where the lobes are clearly of one sign or another and the effect is reasonably large. The case $E_3$ also shows an average that is not symmetric about the horizontal axis, but over a larger distance scale.

In Figure 2 we again consider the difference in probabilities for $\nu_\mu \to \nu_\mu$ for the direct and reverse process, but this time there is ten times as much air as material (Fig. 2a) or ten times as much material as air (Fig. 2b); again the horizontal scale is the amount of air in the same units as in Fig. 1. The effect remains large and systematic.

Finally we remark on an off-diagonal process, $\nu_\mu \to \nu_e$ for definiteness. These processes have order dependence even in the absence of a CP-violating phase $\delta$. Figures 3a-b plot the probability difference for the direct and reverse processes for $\delta = 0$ and $\delta = 1$ respectively. These probabilities certainly differ from one another; however, the standard for the detection of non-zero $\delta$ is different from the diagonal case.

Whether the phenomenon being described here is of eventual use will depend on as-yet unknown features of the neutrino spectrum. Detectable oscillations over managable distances would certainly render our results more relevant. It is nevertheless interesting that there is at least in principle a way to look for CP-violation in the neutrino sector without having to compare neutrino and antineutrino processes.

## Acknowledgements

We would like to thank the Aspen Center for Physics, where much of this work was done. PMF would also like to thank Dominique Schiff and the members of the LPTHE at Université de Paris-Sud for their hospitality. This work is supported in part by the U.S. Department of Energy under grant number DE-FG02-97ER41027.

**Figure Captions**

Figure 1. Probability differences for finding $\nu_\mu$ as a function of time from production as a pure $\nu_\mu$ between the case where the neutrino beam passes through two different layers of equal thickness in one order and the case where the beam passes through the layers in the reverse order. The horizontal axis is the time for passage through (or thickness of) the air layer. Parameter values as described in the text. (a) $E = 100$ MeV; (b) $E = 500$ MeV; (c) $E = 20$ GeV.

Fig 2 As in Fig. 1, except the thickness of the air and the material layer are different. (a) Air thickness 10 times that of material thickness; (b) air thickness 1/10 that of material thickness.

Figure 3. Probability differences for finding $\nu_e$ as a function of time from production as a pure $\nu_\mu$ between the case where the neutrino beam passes through two different layers of equal thickness in one order and the case where the beam passes through the layers in the reverse order. The horizontal axis is the time for passage through (or thickness of) the air layer in units of $10^{13}$ eV$^{-1}$. Parameter values as described in the text. (a) $\delta = 0$; (b) $\delta = 1$.

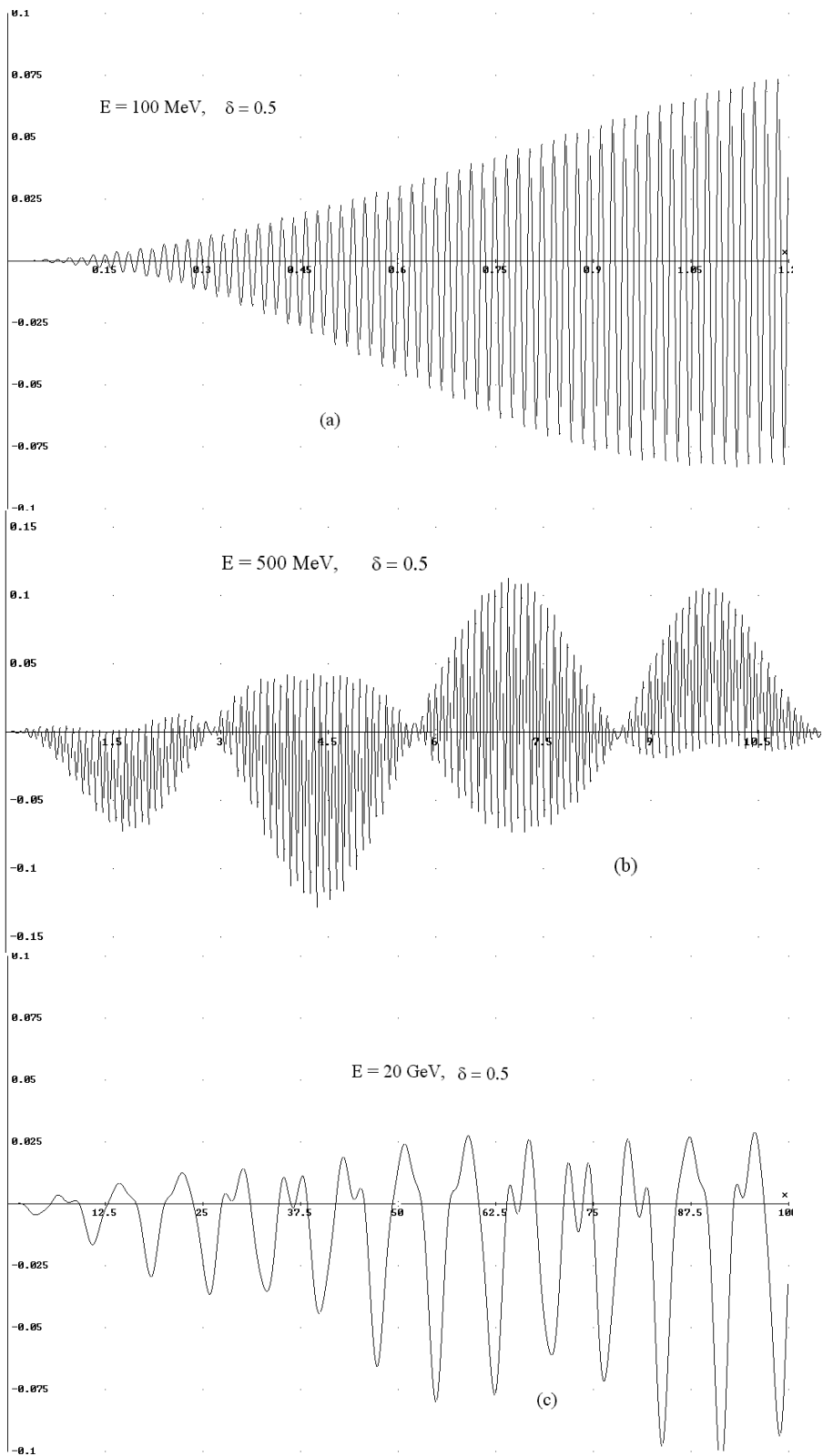

Figure 1

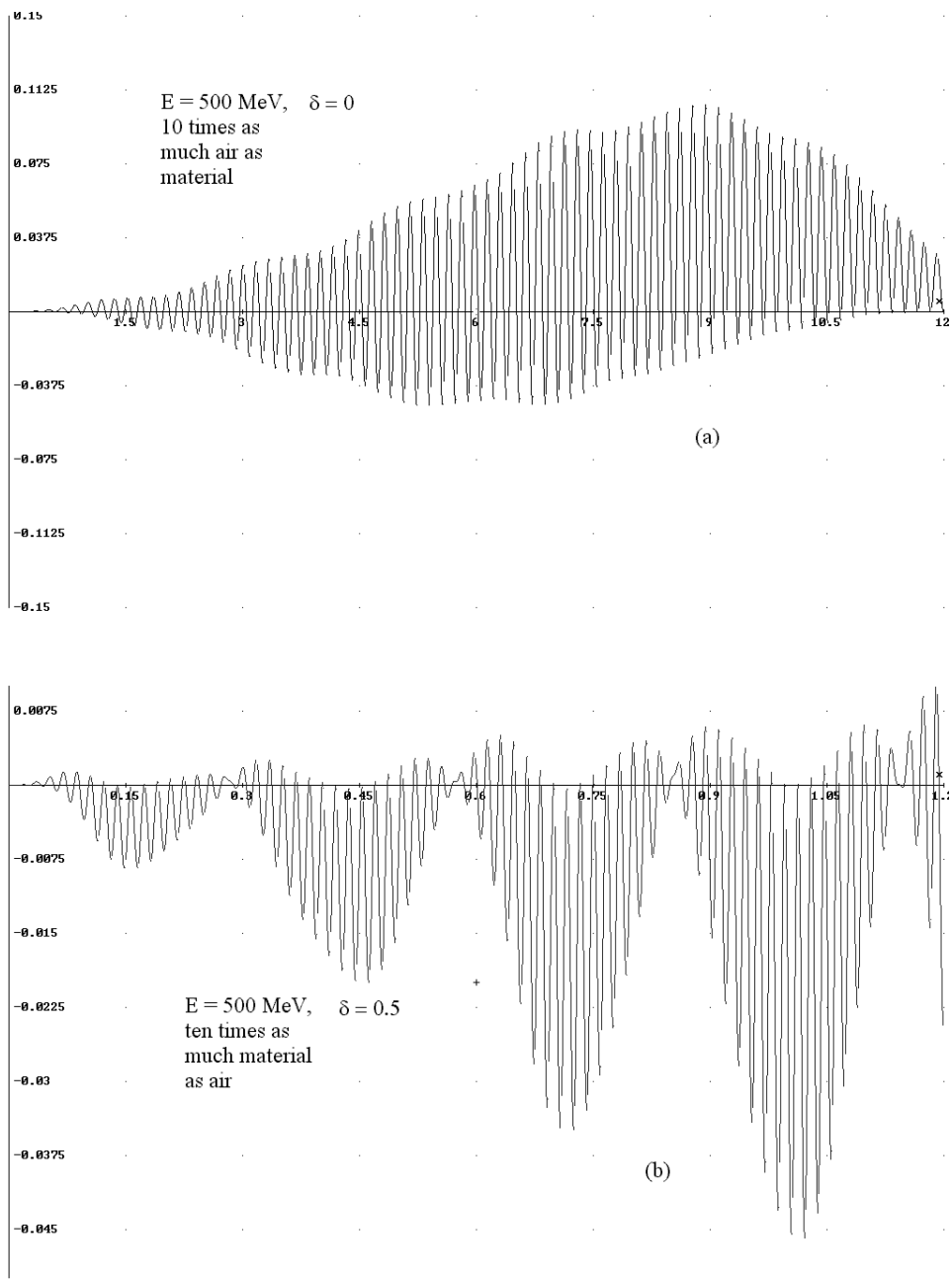

Figure 2

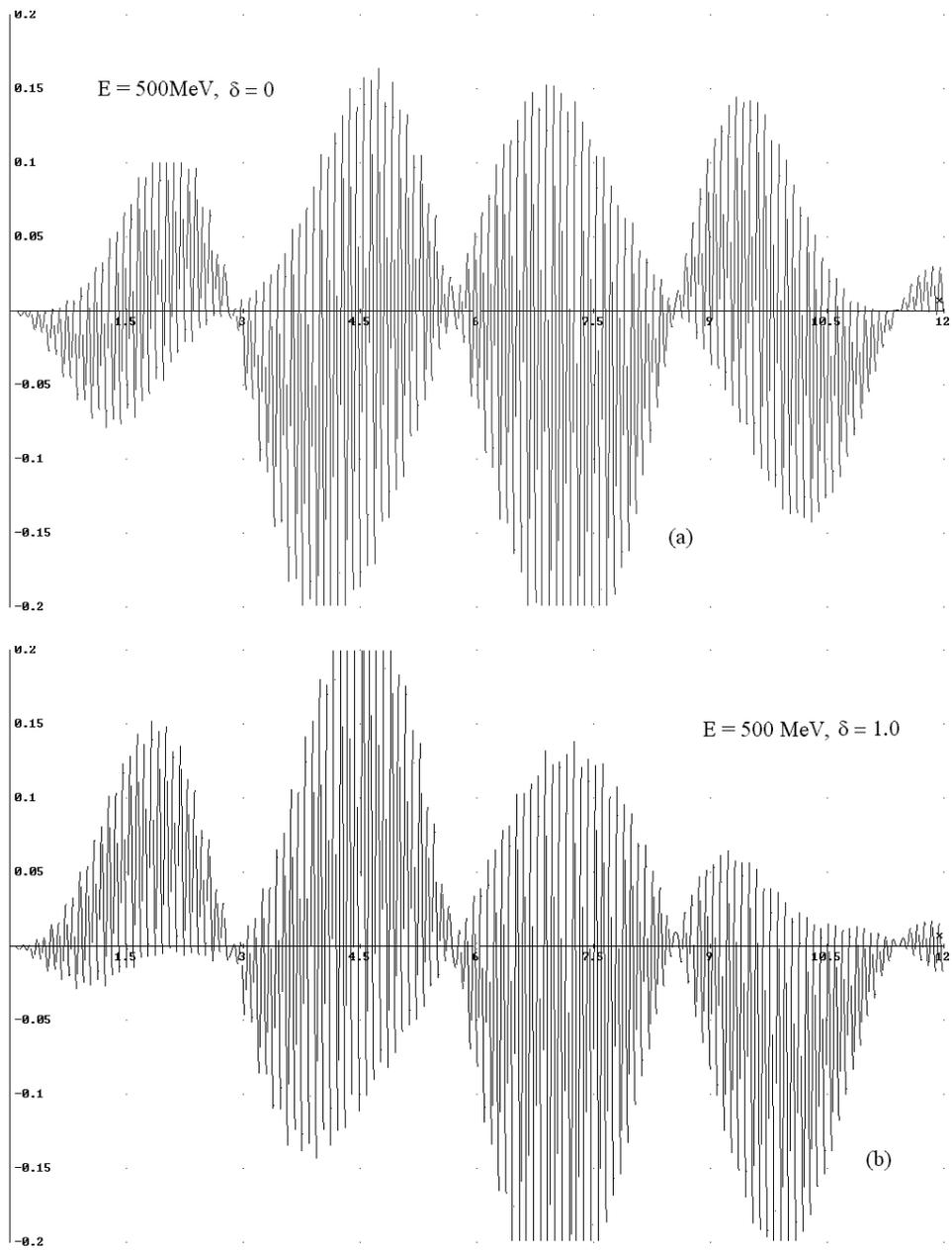

Figure 3